\documentclass[preprint,12pt]{elsarticle}

\usepackage{amssymb}
\usepackage[utf8]{inputenc}
\usepackage[export]{adjustbox}
\usepackage{graphicx}
\usepackage{microtype}
\usepackage{amsmath}
\usepackage{xparse}
\usepackage{xkeyval}
\usepackage{float}
\usepackage{breqn}
\usepackage{xcolor}
\usepackage{tabularx,caption}
\usepackage{booktabs}
\usepackage{multirow}
\usepackage{siunitx}

\journal{}

\makeatletter
\def\ps@pprintTitle{%
 \let\@oddhead\@empty
 \let\@evenhead\@empty
 \def\@oddfoot{\centerline{\thepage}}%
 \let\@evenfoot\@oddfoot}
\makeatother

\begin{document}
\begin{sloppypar}

\begin{frontmatter}



\title{A constitutive model for elastomers tailored by ionic bonds and entanglements}


\author[inst1]{Zhongtong Wang}

\affiliation[inst1]{organization={Sibley School of Mechanical and Aerospace Engineering, Cornell University},
            city={Ithaca},
            postcode={14853}, 
            state={New York},
            country={United States}}

\author[inst2]{Hongyi Cai}
\author[inst1]{Meredith N. Silberstein}

\affiliation[inst2]{organization={Department of Materials Science and Engineering, Cornell University},
            city={Ithaca},
            postcode={14853}, 
            state={New York},
            country={United States}}

\begin{abstract}
 Over the past decade or two, the concept has emerged of using multiple types of weak interactions simultaneously to enhance the mechanical properties of elastomers. These weak interactions include physical entanglements, hydrogen bonds, metal-coordination bonds, dynamic covalent bonds, and ionic bonds. The combination of entanglements and ionic bonding has been minimally explored and is particularly exciting because of the broad application space for polyelectrolytes. In this work, a constitutive model framework is developed to describe the response of elastomers with both ionic bonds and entanglements. We formulate a micromechanical model that couples together chain stretching, ionic bond slipping, and entanglement evolution.  The ionic bonds provide toughness by enabling plastic deformation in comparison to covalently crosslinked material and add strength compared to a linear polymer. Evolution of the entanglement density is taken as a key mechanism that can govern stiffness, toughness, and self-recovery in elastomers. The model is used to match bulk polyelectrolytes with different fractions of ionic components under a variety of loading histories. The variations in material parameters are then used to help understand the relative importance of different governing mechanisms in the bulk polymers.  We show that the theoretical framework can explain our experimental uniaxial tensile experimental results for polyelectrolytes. This model can help to design better material with high stiffness and toughness. We expect that our model can be extended to explain the mechanical behavior of other polyelectrolytes and other soft materials with a wide range of dynamic bonds. 
\end{abstract}



\begin{keyword}
constitutive model \sep polyelectrolytes \sep entanglement
\end{keyword}

\date{}

\end{frontmatter}


\section{Introduction}
\label{sec:sample1}










Polyelectrolytes are polymers possessing a high density of ionizable groups. The combination of polymeric and electrolytic characteristic of polyelectrolytes leads to their use in a range of applications. These materials have been widely applied for purposes such as cell culture scaffolding \citep{drury2003hydrogels,jen1996hydrogels}, drug delivery \citep{hoare2008hydrogels,li2016designing,qiu2001environment}, sensing and actuating \citep{richter2008review,dai2019ultrafast}, soft robotics \citep{maeda2007self}, and wearable electronics \citep{kim2016highly}. Many stimuli-responsive polyelectrolyte elastomers and gels have been developed, including pH-responsive \citep{sui2003controlling,horta2009ph}, electric field responsive \citep{lin2009electric,shang2008electrical}, and light responsive \citep{zhang2014optical}.  Due to these interesting properties, improving the mechanical properties of polyelectrolytes is important. Both dynamics crosslinks \citep{vidavsky2018modulating,vidavsky2020tuning,zhang2020bridging,scheutz2019adaptable,kuang2021magnetic,shen2021nonsteady,miwa2018dynamic,peng2018strong} and entanglements \citep{kong2021control,wendlandt2005non,romano2014influence,kamiyama2022highly,kim2021fracture, bosnjak2021pathways} are widely used in polymeric networks to tailor the mechanical properties.



One common modeling approach for elastomers with dynamic bonds is to start from the statistical mechanical behavior of a single chain, introduce a microphysical model at the network level, and then map the network to the macroscopic deformation. For example, the transient network theory first proposed by Tanaka and Edwards \citep{tanaka1992viscoelastic} and further developed by Vernerey et al. \citep{vernerey2017statistically}, develops a statistical model unifying the evolution of chain/network configuration and kinetics of crosslinks, and is able to describe several time-dependent responses of gels including hysteresis, the Mullins effect, stress stiffening/softening and necking \citep{vernerey2018statistical}. The transient network theory links macroscopic mechanics to the single polymer chain through the chain distribution function. Later, Lamont et al., incorporated a phenomenological criterion of chain damage into transient network theory and considered the bond rupture by adding an enthalpic term into the free energy density \citep{lamont2021rate}. Buche and Silberstein \citep{buche2020statistical, buche2021chain} established a general framework to work from an arbitrary single chain model Hamiltonian to a macroscopic constitutive relations. This theory was demonstrated to capture critical features of dynamically bonded elastomer networks, but struggled with more subtle time dependent mechanisms. While statistical physics-based approaches offer a strong connection of macroscale behavior to molecular mechanisms, obtaining a set of tractable equations often requires making non-physical assumptions for complicated polymer networks.


Alternatively, constitutive models for elastomers with dynamic bonds can be formulated from the macroscale and utilize intact crosslink fraction as a state variable on which the stress depends. For example, Hui and Long \citep{hui2012constitutive} keep track of the time evolution of the physical bonds which are assumed to be only controlled by the maximum deformation and established a theoretical framework of an ionically crosslinked triblock copolymer gel. Later, Hui and coworkers extended the framework and captured two essential mechanisms: relaxation due to chain detachment and self-healing due to chain reattachment for a PVA gel by assuming that the chain dissociation process is time dependent and insensitive to deformation \citep{long2014time}. Using a similar approach, Venkata et al \citep{venkata2021constitutive} established a framework to capture the time-dependent mechanical behavior of pure physical polyampholyte (PA) gels by using two types of bond kinetics since pure PA gels (without a chemical crosslinker) have a phase separated structure. Lin et al. \citep{lin2020constitutive} developed a constitutive model to capture the time-dependent stress response, relaxation, hysteresis and self-healing behavior of a physical hydrogel by considering the viscous stretch recovery mechanism of conformational change, rather than just considering the viscous stretch evolving with the applied stress. Taking a different approach, Yu and coworkers tracked the evolution of the number of microscopic dynamic crosslinks and described the constitutive models for self healing behavior \citep{yu2018mechanics} in which the healing process occurs from a coupling of polymer chain diffusion and dynamic bond binding. In this paper we will follow a state variable-based macroscale constitutive modeling approach. However, in contrast to much of the prior work on dynamic bonds in elastomers and gels, our dynamic bonds will be assumed to reform instantly (more similar to associative vitrimers at the bond scale\citep{krishnakumar2020vitrimers}, but distinct at the network scale), providing toughness by acting as a plasticity mechanism rather than a stiffness reduction and recovery mechanism.

Chain entanglement impedes the movement of molecular segments and influences polymer rheology, morphology, and mechanical properties \citep{huang2000influence, bartczak2005influence, luo2017molecular, pawlak2018cavitation, pawlak2019entanglements}. The highly entangled equilibrium state of polymers is well described by the tube model, in which each polymer chain is essentially confined in a tube-like region due to topological constraints formed by other chains, and diffuses along a tube of its contour length\citep{edwards1988tube}. Expanding the tube model, Rubinstein and Panyukov introduced the affine length parameter to the model, separating the deformation of the elastomers into affine deformation and nonaffine deformation. The nonlinear elasticity of the elastomers is induced by nonaffine deformation \citep{rubinstein1997nonaffine}. By replacing the Gaussian distribution with the Langevin distribution, Davidson and Goulbourne presented a nonaffine network model to capture the strain hardening \citep{davidson2013nonaffine}. Following the idea of Rubinstein and Panyukov regarding the slip-tube model \citep{rubinstein2002elasticity} and considering the motion of the polymer chain, Xiang et al \citep{xiang2018general} decomposed the free energy into the unentangled crosslinked network and the entanglements between the chains. Silberstein et. al. implemented a rate and stretch history dependent entanglement length evolution for linear elastomers to account for inelasticity \citep{silberstein2014modeling}.  Despite these works modeling entanglement constraints, it remains elusive how to model the dynamic release and reattachment entanglements under large deformation. In contrast to much of the previous work adding the entanglement constraints to network deformation, we will mechanistically model both entanglement loss and recovery.

In order to tailor the mechanical properties of polyelectrolytes, it is necessary to understand the mechanisms that govern those properties. To that end, we develop here a theoretical framework for the ionically bonded and entangled polymers. We use prior experimental work from our group \citep{cai2022highly} to inspire and validate the constitutive model. This model must include mechanisms to explain three important features observed experimentally: the first is the reduction of elastic modulus, the second is the plastic deformation, and the third is the self-recovery over time. We utilize a reference system that consists of polymer chains with only one type of charge (and therefore containing no ionic bonds) to isolate the effect of ionic crosslinks on the overall bulk response. Entanglement evolution with shear stress and diffusion is considered for the modulus reduction and self-recovery. 

The plan of the paper is as follows. In section \ref{sec:ex}, we show key experimental behavior of ionically bonded and entangled bulk polyelectrolytes. In section \ref{sec:model}, based on the experimental observations and underlying physical mechanisms, a spring-dashpot model is established to capture the main features of the stress-strain behavior. In section \ref{sec:result}, we show the constitutive model behavior dependence on key parameters. We then compare our simulation results with the experimental results for  the single component polymers and the complex with ionic bonding. We then use those fitting parameters to predict the experimental results for polyelectrolyte polymers with different fractions of ionic components.

\section{Experiments} \label{sec:ex}
\label{sec:sample2}






The polyelectrolytes used in this work to inspire and provide a fit for the model are ethyl acrylate (EA) copolymers synthesized by free radical polymerization. The synthesis procedure and extensive experimental characterization are published in the authors' previous work \citep{cai2022highly}. We use [2-(acryloyloxy)ethyl]trimethylammonium chloride (ATMAC) for the cationic copolymers and 3-sulfopropyl acrylate potassium (SPAP) for the anionic copolymers. These polymer chains are therefore either positive or negative. For simplicity, each copolymer is named by its neutral monomer identity, target ratio of neutral to charged monomer, and whether it has positive or negative charge. For example,  the cationic copolymer with an ingredient monomer ratio EA:ATMAC = 20:1 is EA 20+. The anionic copolymer with an ingredient monomer ratio EA:SPAP = 20:1 is EA20-. EA20+/- was produced by mixing EA20+ and EA20- and dialyzing out the small counter ions. The polymers were solvent casted into 1 mm thick sheets of material. The complex EA20+/- has a $T_g$ of -5.10 °C, in between that of the positive component EA20+ (-5.95 °C) and the negative component EA20- (-3.07 °C). The mechanical behavior of the polyelectrolytes in this work has been thoroughly investigated including monotonic stretching at different loading rates, cyclic deformation, and stress relaxation. All the experiments were performed on a ZwickRoell Z010 universal testing machine with a 20 N load cell at room temperature, using crosshead displacement control and constant engineering strain rates. The specimens were cut from polyelectrolyte sheets into rectangular shapes with a width of 4 mm, and gripped with an initial gauge length of 20 mm.

\begin{figure}[H]
    \centering
    \includegraphics[width=1.2\textwidth,center]{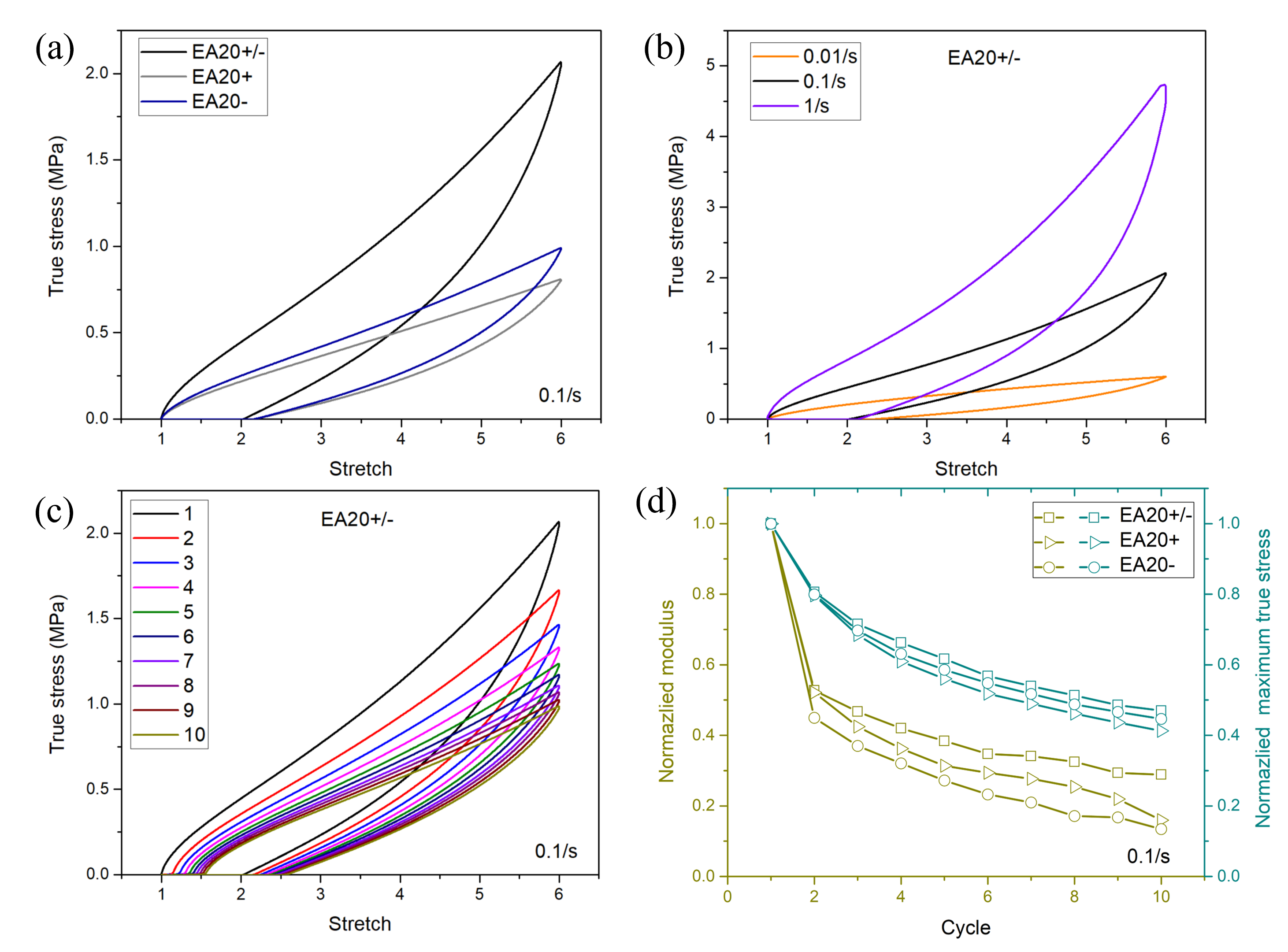}
    \caption{Uniaxial tensile experimental results. (a) A single cycle at a strain rate of 0.1/s for EA20+/-, EA20+ and EA20-. (b) Rate dependent experimental results for EA20+/- at three different strain rates. (c) Cyclic loading for EA 20+/- at a strain rate of 0.1/s with a wait time of 10 min between cycles. (d) The normalized modulus (left axis) and normalized maximum true stress (right axis) obtained from cyclic tensile loading tests at a 0.1/s strain rate for EA20+/-, EA20+ and EA20-.}
    \label{fig:expsum}
    \end{figure}

Figure \ref{fig:expsum}a shows the true stress versus stretch curve of the single component (EA20+ and EA20-) and complex (EA20+/-) at a strain rate of 0.1/s deformed to a stretch of 6 and then unloaded. The true stress-stretch curve shows substantial hysteresis and has a similar shape for the individual components and the complex. The complex has approximately twice the initial elastic modulus as the individual components. Figure \ref{fig:expsum}b shows the true stress-stretch curves for EA20+/- at three different strain rates ranging from 0.01/s to 1/s. In addition to the large hysteresis, EA20+/- also shows a significant strain rate sensitivity. Because the complex is ionically bonded and entangled, the rate dependence likely arises from both ionic bond exchange and entanglement slip. 

Material behavior under multiple cycles to the same strain was also investigated. For each cycle, the specimens were first deformed to a stretch of 6 at a strain rate of 0.1/s and retracted down to the unstressed state at the same loading rate. While the crosshead returns to its initial position, the thin specimen reaches zero tension before that point, and bends rather than axially compressing - in our simulations we will approximate this portion of the loading history as a zero stress condition. Figure \ref{fig:expsum}c shows the results of 10 cycles with a hold time of 10 min between cycles. Upon stretching during the first cycle, the polyelectrolytes exhibit a large hysteresis loop. As the sample undergoes more cycles, softening of the stress can be observed. It is clear that the slope of the second loading is much higher than the first unloading curve, indicating that the stiffness of polyelectrolyte shows self-recovery over time. Figure \ref{fig:expsum}d shows that the normalized elastic modulus and normalized maximum true stress decay monotonically with increasing cycle number. The individual copolymer and complex polyelectrolytes show similar behaviors in terms of modulus and maximum true stress reduction. Because there are no ionic crosslinks in the individual copolymers, these crosslinks cannot be responsible for the large modulus drop and stress decrease. We therefore hypothesize that upon stretch during the first cycle, entanglement evolution leads to the softening of the stress. Entanglement recovery then occurs during unloading and the 10 minute waiting time. As the sample undergoes more cycles, the elastic modulus and the maximum true stress continue to decrease. Since the modulus reduction of the complex is similar to the individual copolymer (Figure \ref{fig:expsum}d), we assume that the entanglement evolution dominates the modulus evolution and that ionic crosslinks will break and reform, with the total number of the ionic crosslinks not changing. This concept is also logical from a thermodynamics perspective, since an unpaired ionic charge would have an electric field associated with it.

\section{Model framework}  \label{sec:model}

\subsection{Overview of the material system}

\begin{figure}[H]
    \centering
    \includegraphics[width=1.2\textwidth,center]{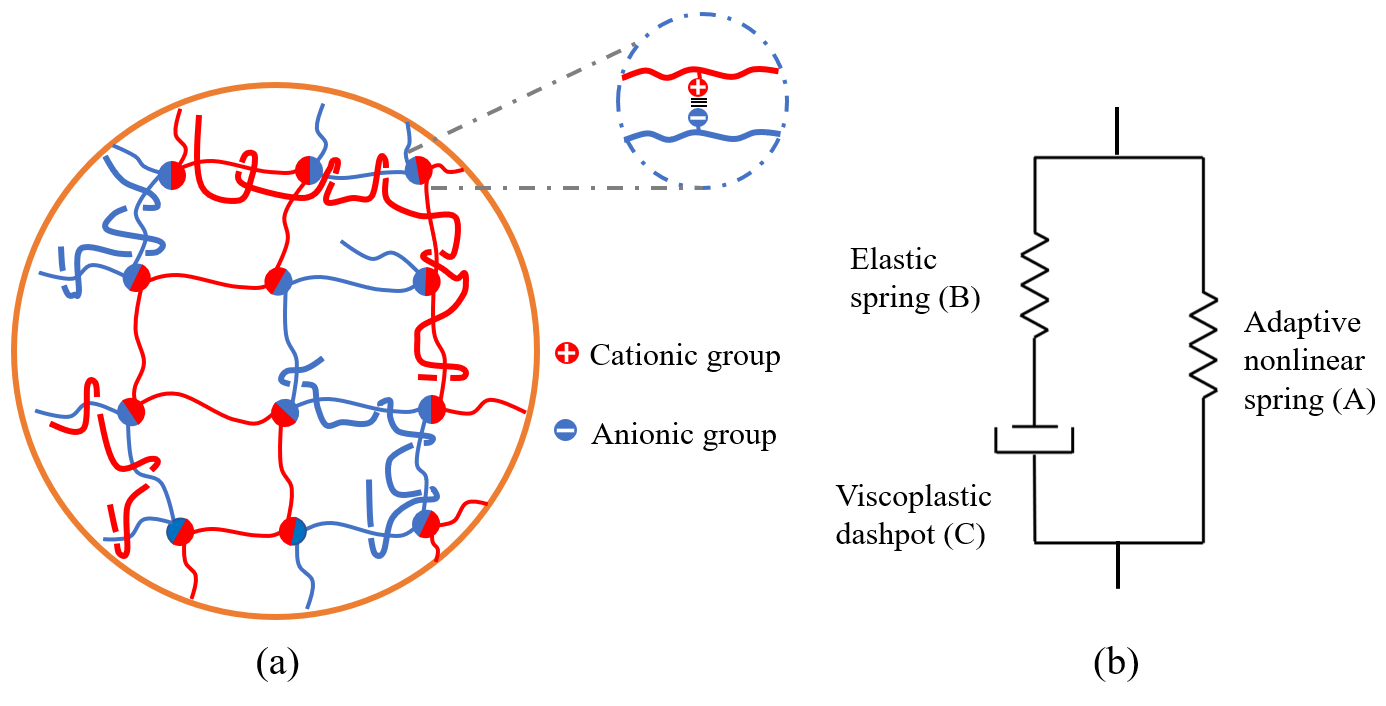}    
    \caption{(a) Schematic showing the ionic crosslinks and entanglements of the complex assembled from two oppositely charged copolymers. (b) A spring dashpot representation of the constitutive model.}
    \label{fig:schematic}
    \end{figure}

The polyelectrolytes we are modeling here consist of long, flexible (above $T_g$), chains that interact with each other through ionic bonds and physical entanglements as shown in Figure \ref{fig:schematic}a. In order to isolate the mechanism of ionic crosslinking on the overall bulk response, we utilize a reference system, the material consisting of polymer chains of only one type of charge, which is then balanced by small molecule counter ions. This reference system therefore has no ionic crosslinking. As is clear from the experimental data and expected from literature, entanglements can significantly contribute to the overall stress response of an elastomer.  It is also clear from both literature and our experiments that ionic crosslinks add stiffness and strength, and influence inelastic recovery. In our bulk polelectrolyte model, we therefore consider polymer chain stretching, ionic bond sliding, and entanglement slip and recovery.

The 3D, large deformation model consists of three constitutive elements, illustrated schematically by the spring-dashpot representation in Figure \ref{fig:schematic}b. An adaptive nonlinear spring A is in parallel with the element B-C, which consists of a linear spring (B) in series with a viscoplastic dashpot (C).

In the following, superscripts $A,B$ and $C$ denote the variables corresponding to the elements shown in Figure \ref{fig:schematic}b. Due to the parallel arrangement of these components, the overall deformation gradient of the polyelectrolytes $\textbf{F}$ can be expressed as 

\begin{align}
    &\textbf{F} = \textbf{F}^A = \textbf{F}^{BC}  
\end{align}

\noindent where $\textbf{F}^A$  and $\textbf{F}^{BC}$ are the deformation gradients acting on the nonlinear spring (A) and  the linear elastic-viscoplastic component (BC).  The total true stress $\textbf{T}$ is given by

\begin{align}
    & \textbf{T} = \textbf{T}^A + \textbf{T}^{BC}
\end{align}

\noindent where $ \textbf{T}^A$ and $\textbf{T}^{BC}$ are the contribution to the true stress originating from the nonlinear spring and the linear elastic-viscoplastic part, respectively. Due to the series configuration of the BC element, the true stress of spring element B and dashpot element C are equal, namely, 

\begin{align}
    & \textbf{T}^{BC} = \textbf{T}^B = \textbf{T}^{C}
\end{align}

\noindent where $ \textbf{T}^B$ and $\textbf{T}^C$ are the true stress of the spring element B and dashpot element C, respectively. Constitutive equations for each component are described below.

\subsection{Polymer network behavior}

The polymer network behavior will be captured by a nonlinear spring, which expresses both the stretching of chain segment between entanglements and evolution of the density of these entanglements.

\subsubsection{Chain stretching}

In order to capture chain stretch micromechanically, we use the Langevin chain-based Arruda-Boyce model \citep{arruda1993three}. The true stress in the network mechanism is expressed as:

\begin{align}
    & \textbf{T}^A =  \frac{ nkT \sqrt{N}\bar{\textbf{B}}_s}{3J \Lambda} \mathcal{L}^{-1}\left(\frac{\Lambda}{\sqrt{N}}\right)
\end{align}

\noindent where $n$ is the number density of entanglements, $k$ is Boltzmann's constant, $T$ is absolute temperature, and $N$ is the number of Kuhn segments between physical entanglements and sets the limiting chain extensibility. $J = \det \textbf{F}$ is the relative volume change.  $\bar{\textbf{B}}$ is the isochoric left Cauchy-Green tensor, $ \bar{\textbf{B}}=\bar{\textbf{F}} (\bar{\textbf{F}})^T $ and $ \bar{\textbf{B}}_s =\bar{\textbf{B}} -tr(\bar{\textbf{B}})\textbf{I}/3$  is the deviatoric part of $\bar{\textbf{B}}$. $\bar{\textbf{F}} = (J)^{-1/3}\textbf{F}$ is the isochoric part of $\textbf{F}$. $\mathcal{L}^{-1}(\cdot)$ is the inverse Langevin function, $\mathcal{L}^{-1}(x) = \coth{x} - x^{-1}$, which is computed using the Pad\'e approximation \citep{jedynak2015approximation}. Subject to a stress, the network undergoes affine deformation with principal stretches of $\lambda_1, \lambda_2$, and $\lambda_3$. Therefore, the stretch of each representative chain is 
\begin{align}
    & \Lambda = \sqrt{\frac{\lambda_1^2 +\lambda_2^2 +\lambda_3^2 }{3}} 
\end{align}

\noindent $N$ and $n$ evolve as entanglements slip and recover as described in the following section.

\subsubsection{Entanglement evolution}

Evolution of the entanglement density is taken as a key mechanism that can govern stiffness, toughness, and self-recovery in elastomers. During loading, the number of Kuhn segments per chain segment increases due to the release of entanglements, so that the limiting chain extensibility increases. When stress is reduced, the number of effective segments per chain can recover due to the reattachment of entanglements, so that the limiting chain extensibility decreases. Mass conservation requires that,

    \begin{align}  
    & n_0 N_0 = n N 
    \end{align}
    
\noindent where $n_0$ is the initial number density of entanglements and $N_0$ is the initial number of segments between entanglements.

Entangled polymers are well described by the reptation model of De Gennes and Edwards \citep{edwards1988tube,de1979scaling}. The basic idea of the reptation model is that the polymer chain is constrained by the polymer matrix so it can only reptate along a primitive tube.  The curvilinear motion of the polymer chain is characterized by the Rouse friction model \citep{doi1988theory} with curvilinear diffusion coefficient ($D$): 

    \begin{align}  
    & D = \frac{k T}{ N_0 \eta} 
    \end{align}
    
\noindent where $\eta$ is the Rouse friction coefficient per Kuhn segment. The primitive tube length $L_c$ is taken as $Nb^2/a_c$, where $a_c$ is a step length of the primitive chain and  $b$ is the length of the Kuhn segment. In general, the polymer chain diffuses along its contour length in a tube of diameter much smaller than the length of the chain following a primitive (curvilinear) path \citep{edwards1988tube}. We assume the step length is equal to the length of a Kuhn segment, that is $a_c = b$ \citep{yu2018mechanics}. In the following, we have a two coordinate system, s and y, where s denotes the curvilinear path along the chains and y denotes the horizontal projection. The conversion of distances in the two systems is given by $y = \sqrt{sb}$ \citep{yu2018mechanics,edwards1988tube}. The entangled polymers move along the primitive path. Here, we use the term ``overlap distance" to denote the length along the primitive path a polymer needs to reptate to release the entangled polymers. For simplicity, we assume that the fully entangled polymer will be released from entanglement when it moves half of the primitive tube length, but this decision effectively shifts fitting parameters. We will use ``overlap end-to-end distance" to denote the horizontal projection of overlap distance. Here, $L = \sqrt{N}b$ expresses the end to end distance for the entangled polymer. Therefore,  the length for disentanglement along the horizontal projection is $L/2$ which is also used for the re-entanglement process.

Since the polymer chains are randomly entangled with each other, the overlap distance for the entangled polymer in polyelectrolytes has a distribution. We treat this distribution as $m$ populations of entangled networks. The horizontal projection of the overlap distance $s_i$ is distributed normally about an average value, where $1\leq i \leq m$. The chain probability density of overlap distance can then be expressed by the normal distribution: 

\begin{align}
    & P_i(s_i) = \frac{1}{\delta \sqrt{2\pi}}\exp\left[-\frac{\left(s_i-\psi\right)^2}{2\delta^2}\right]
\end{align}

\noindent where $\psi$ and $\delta$ denote the average value and the distribution width, respectively. 

The evolution of entanglement involves both loss of entanglement density under stress and recovery of entanglement density over time. As a polymer network is stretched, shear stress in the network drives the relative chain slippage and disentanglement. Over time, diffusion enables polymer chain re-entanglement. Inspiring by the Ree-Eyring's theory \citep{ree1955theory}, we modify the reptation model to include shear stress driven loss of entanglement, coupled with stress independent recovery. The equations for entanglement evolution are expressed as

 \begin{align} 
    & \frac{\partial n_i^a\left(t,s_i\right)}{\partial t}= D \frac{\partial^2 n_i^a \left(t,s_i\right)}{\partial s_i^2} - k_r  n_i^a\left(t,s_i\right)  \label{eq:entangle_law}\\
    & k_r = k_0 \exp \left(-\frac{\Delta G_t}{kT}\right) \sinh \left(\frac{\Delta G_t}{kT} \frac{\tau_t}{f_t}\right)\\
    & \tau_t = \sqrt{\frac{\textbf{T}_s:\textbf{T}_s}{2}}
 \end{align}

\noindent where $n_i^a$ is the density of the ith population of entangled chains along the curvilinear coordinate s at time t, $k_r$ is the reaction rate of shear stress driven chain diffusion, $k_0$ is a prefactor for the reaction rate, $\Delta G_t$ is activation energy for the shear stress driven loss of entanglements, and $f_t$ is the deformation resistance parameter for entanglement. $\textbf{T}_s$ is the deviatoric part of the total stress. 

To solve equation \ref{eq:entangle_law}, we need initial and boundary conditions. In the initial state, entanglement density is at the stress free equilibrium and therefore at it's maximum. The initial normalized number of entanglements is set to be 1. The normalized number of entanglements will stay at 1 on the boundary, since all entangled polymers have a tendency to remain active. The horizontal projection of the overlap distance is $L_i/2$ where $i$ indicates the population number and $y_i = \sqrt{s_i b}$. Therefore, the positions $y_i = 0$ and $y_i = L_i/2 $ correspond to $s_i = 0$ and $s_i = L_i^2/4b$. Therefore, the boundary and initial conditions are:
    
    \begin{align}  
    & n_i^a\left(t=0,s_i\right) = 1  \label{eq:inital}\\
    & n_i^a\left(t,s_i = 0\right) = 1 \label{eq:boundary1} \\
    & n_i^a\left(t,s_i = \frac{L_i^2}{4b}\right) = 1 \label{eq:boundary2} 
    \end{align}

\noindent The average density of entanglements $\alpha$ for the total system is calculated by:

    \begin{align}  
    & \alpha\left(t\right) = \sum_{i=1}^{m} P_i \frac{4b}{L_i^2} \int_{0}^{L_i^2/4b} n_i^a\left(t,s\right) \,ds \
    \end{align}
\noindent We then obtain the number of entanglements that remains at time t by: 

    \begin{align}  
    & n = n_0   \alpha \left(t\right)\
    \end{align}

\subsection{Linear elastic-viscoplastic element}
  
The linear elastic-viscoplastic leg of the model captures stretch of the chain segments between ionic crosslinks and the sliding of chains relative to each other as ionic crosslinks break and reform. The total deformation gradient is decomposed into the elastic and plastic components. 

\begin{align}
    &\textbf{F}^{BC} = \textbf{F}^{B}\textbf{F}^{C} 
\end{align}

\noindent where $\textbf{F}^B$ and $\textbf{F}^C$ are the deformation gradients of the spring element B and dashpot element C respectively. Plastic deformation is assumed to be incompressible. 

\begin{align}
    &J^{C} = det\left(\textbf{F}^{C}\right) = 1 
\end{align}

\noindent where $J^{C}$ is the relative volume change for viscoplastic element (C). 

The true stress of the elastic viscoplastic element can be calculated as 

\begin{align}
    &\textbf{T}^{BC} = \frac{1}{J} \textbf{R}^{B} \textbf{M}^{BC} \left(\textbf{R}^{B}\right)^T 
\end{align}

\noindent where $\textbf{M}^{BC}$ is the Mandel stress and $\textbf{R}^{B}$ is the rotation tensor that can be obtained through the polar decomposition of the deformation gradient:

\begin{align}
    &\textbf{F}^{B} = \textbf{R}^{B}\textbf{U}^{B}
\end{align}

\noindent $\textbf{U}^{B}$ is the elastic stretch tensor of spring B. 

The Mandel stress $\textbf{M}^{BC}$ is prescribed as

\begin{align}
    &\textbf{M}^{BC} = 2 \mu \ln \textbf{U}^B + \left(\kappa - \frac{2}{3} \mu\right) tr\left(\ln \textbf{U}^B\right)\textbf{I} 
\end{align}

\noindent where $\mu$ and $\kappa$ are the shear and bulk modulus respectively, and $\textbf{I}$ is the identity tensor.  

The evolution of plastic flow is given by
\begin{align}
    &\dot{\textbf{F}^C} = \textbf{D}^{C}\textbf{F}^{C} 
\end{align}

\noindent with the viscous stretching prescribed as: 
\begin{align}
    &\textbf{D}^{C} = \frac{\dot{\gamma}^C \textbf{M}_s^{BC}}{2\tau} \label{eq:viscous_stretch}  
\end{align}

\noindent where $\textbf{M}_s^{BC} = \textbf{M}^{BC} - \frac{1}{3} tr\left(\textbf{M}^{BC}\right)\textbf{I} $ is the deviatoric part of the Mandel stress. The plastic shear strain rate $\dot{\gamma}^C$ is prescribed by a modified Ree-Eyring viscoplastic model \citep{ree1955theory}. The shear stress is the driving force behind the activation process which allows polymer chains to transition to a new, plastically deformed configuration. The model yields the following equation:

\begin{align} 
    &\dot{\gamma}^C = \dot{\gamma_0} \exp \left(-\frac{\Delta G}{kT}\right) \sinh\left(\frac{\Delta G}{kT} \frac{\tau}{f}\right)  \label{eq:plastic_flow} 
\end{align}

\noindent where $\dot{\gamma_0}$ is the prefactor proportional to attempt frequency, $\Delta G$ is the activation energy for plastic deformation, and $f$ is the deformation resistance. The equivalent shear stress $\tau$ is given by

\begin{align} 
    &\tau = \sqrt{\frac{\textbf{M}_s^{BC}:\textbf{M}_s^{BC}}{2}} 
\end{align}

\section{Results and Discussion} \label{sec:result}


Important features built into this model include the rate of entanglement loss depending on total shear stress, the rate of entanglement increase depending on the diffusion constant, and the rate of plastic deformation, depending on the shear stress for the viscoplastic part. Here,  we first present the tension simulation results at a stretch rate of 0.1/s to highlight those important features. Unless otherwise specified, all parameters regarding the parameter study are as follows: $\mu = 10 kPa, \kappa = 800 kPa, f = 59552 Pa, \dot{\gamma_0}  = 1/s, \Delta G = kT, nkT = 20.7kPa, N_0 = 80, \psi = 0.2,\delta = 0.5, D = 5.2208 \times 10^{-17} m^2/s, b = 3.85nm, \Delta G_t = kT, f_t = 5e6 Pa, k_0 = 1/s$.

\begin{figure}[H]
    \centering
    \includegraphics[width=1.2\textwidth,center]{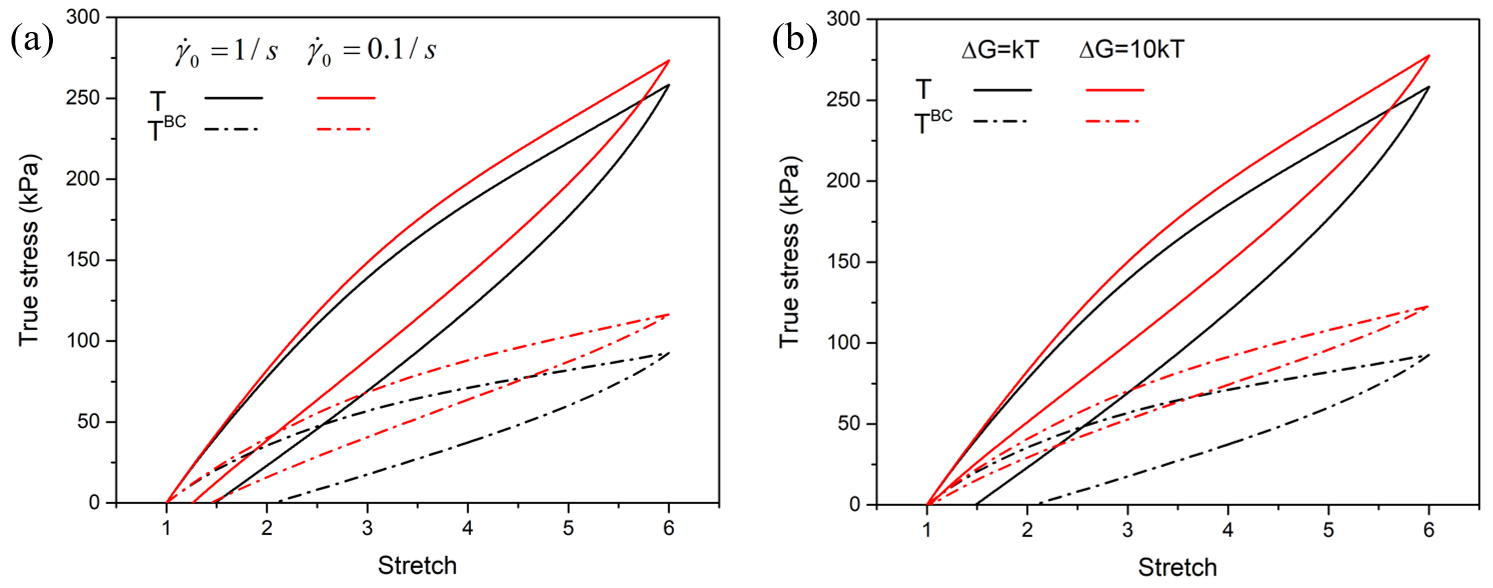}
    \caption{The true stress-stretch curves of the constitutive model at a stretch rate of 0.1/s with different parameters. The contribution of the response from the BC leg of the model is also shown to highlight the contribution of the elastic-viscoplastic component. (a) Varying the prefactor  $\dot{\gamma_0}$ from $ 0.1/s$ to $1/s$. (b) Varying the activation energy $\Delta G$ from $kT$ to $10 kT$}
    \label{fig:parameter_plastic}
    \end{figure}

In the theoretical framework, plastic flow (equation \ref{eq:plastic_flow}) is a thermally activated process that includes two independent parameters, the prefactor $\dot{\gamma_0}$ and the activation energy $\Delta G$.  When increasing the prefactor $\dot{\gamma_0}$ from $ 0.1/s$ to $1/s$, the plastic flow rate increases at the same level of stretch. This increased plastic flow leads to an increased residual strain (Figure \ref{fig:parameter_plastic}a). Correspondingly, the ratio of the stretch of the viscoplastic element increases relative to the stretch of the elastic spring increases with increasing $\dot{\gamma_0}$; therefore, the true stress of the elastic-viscoplastic leg $T^{BC}$ is decreased.   The true stress-stretch behavior dependence on the activation energy $\Delta G$ is similar to, though opposite in sign, the parameter $\dot{\gamma_0}$ (Figure \ref{fig:parameter_plastic}b). At a small energy barrier, the plastic flow readily overcomes the energy barrier and  shows a large residual strain and a small true stress. When the energy barrier is increased by a factor of 10, the residual strain is drastically reduced and the total true stress is larger. Since the decreased plastic flow leads to an increased stretch of the elastic spring, the true stress of the elastic viscoplastic $T^{BC}$ increases. 

\begin{figure}[H]
    \centering
    \includegraphics[width=1.2\textwidth,center]{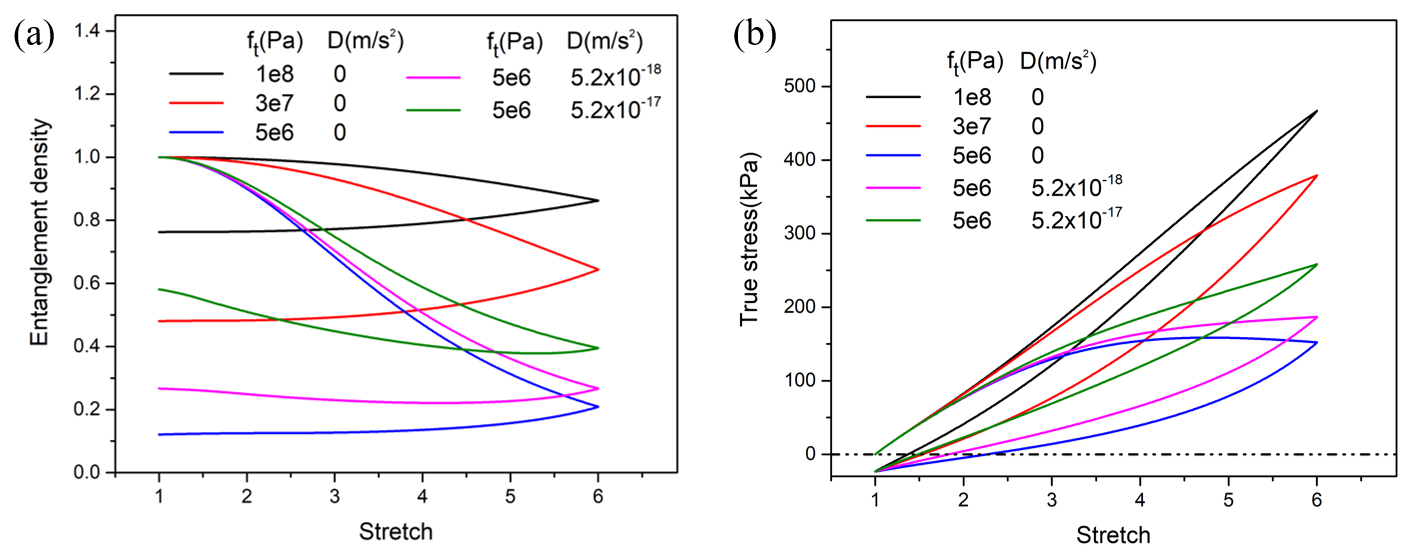}
    \caption{The simulation result for the different parameters $f_{t}$ and  diffusion constant $D$  (a) the density of entanglement; (b) the true stress-stretch curve.}
    \label{fig:parameter_entanglement}
    \end{figure}

Entanglement slip and recovery, captured by equation \ref{eq:entangle_law}, is the second key time dependent mechanism for this model. We look first here at the shear stress driven loss of entanglements.  In order to avoid confusion with the recovery process, we set the diffusion constant $D = 0$. Therefore the shear stress driven loss of entanglements will dominate the loading and unloading process. It is clear that the density of entanglements will always decrease during the uniaxial tension simulation in Figure \ref{fig:parameter_entanglement}a. The deformation resistance parameter for shear stress driven loss of entanglements is similar to the parameter in the yield equation \ref{eq:plastic_flow}. When we decrease deformation resistance for entanglement $f_{t}$, the stress driven loss of entanglement can happen more easily, leading to greater reduction in the density of entanglements and therefore a smaller true stress (Figure \ref{fig:parameter_entanglement}b). 

Next, we increase the diffusion constant $D$ from $0$ to $ 5.2 \times 10^{-17} m^2/s$ (Figure \ref{fig:parameter_entanglement}). As $D$ is increased, the balance between entanglement loss and recovery during loading is shifted, leading to less overall loss of entanglement density, and consequently a return of the stress-strain response towards the no entanglement evolution case. An increase in $D$ also leads to more recovery of entanglements during the unloading process, when the shear driven loss of entanglements is relatively minor. Collectively, this leads to reduced hysteresis and reduces changes in stiffness as $D$ increases. Although not shown here, if the simulated specimen is allowed to rest under no load, the normalized entanglement density will continue to recover towards 1, leading to a larger reloading slope than unloading slope. 


\begin{figure}[H]
    \centering
    \includegraphics[width=1.2\textwidth,center]{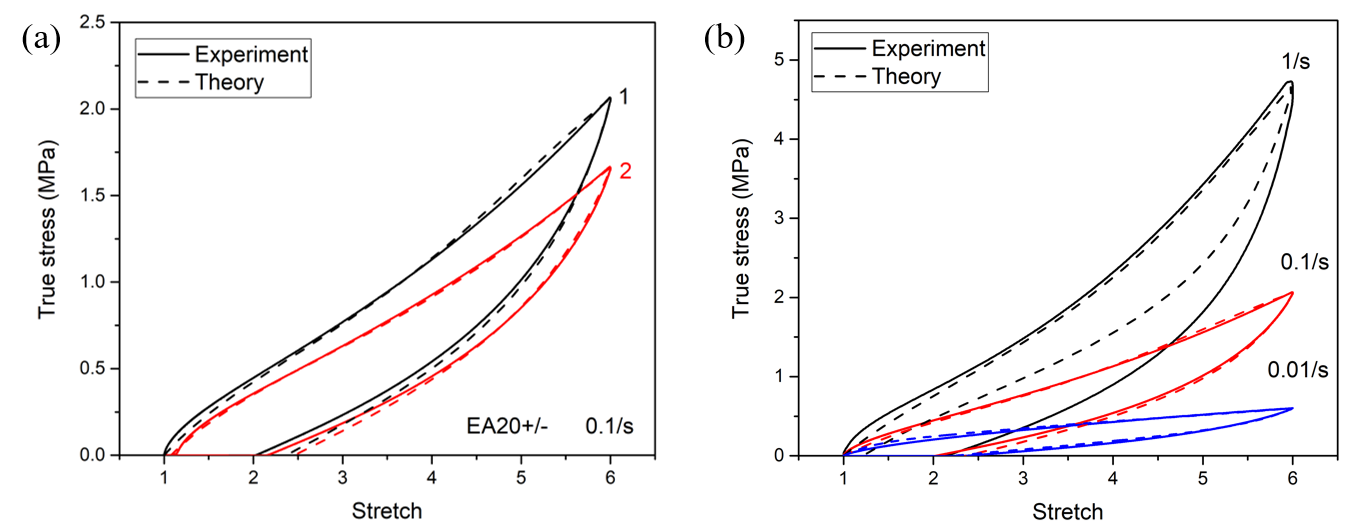}
    \caption{(a) The fitting result for uniaxial tension of 1st and 2nd cycle of EA 20+/- at a strain rate of 0.1/s.  (b) The fitting result for rate dependent experiment for EA 20+/- over a single cycle. Fitting parameters of EA20+/- are given in Table \ref{tab:parameter}.}
    \label{fig:uniaxial_rate}
    \end{figure}

Next, we compare the model with the experimental results. In order to get fitting parameters for the theory, we use  the first and second cycle and the rate dependent experiments on EA20+/- as shown in Figure \ref{fig:uniaxial_rate}. A genetic algorithm is applied here to obtain the fitting parameters. The fitting parameters are given in Table \ref{tab:parameter}. The model captures the true stress stretch curves well for the first and second cycle, especially in terms of hysteresis and residual strain.  For the rate dependent experiment in Figure \ref{fig:uniaxial_rate}b,  EA20+/- shows a significant strain rate sensitivity of the stiffness and hardening which is caused by the plastic flow and entanglement evolution. The fitting result for the model shows a reasonable agreement with the rate dependent experiments in Figure \ref{fig:uniaxial_rate}b, simultaneously capturing the stress level for different rates and hysteresis for low rates.  The highest rate result has a larger error between simulation and experiment. The model can capture the stress level but cannot capture the large hysteresis for highest rate. The discrepancy can likely be attributed to the functional form of the ionic sliding mechanism - perhaps at this faster rate the ionic bond reforming is not effectively instantaneous.





\begin{center}
\captionof{table}{Fitting parameters used in this paper} \label{tab:parameter}
\begin{tabular}{llllll} 
\toprule
{Parameter} & {EA20+/-} & {EA20+} & {EA20-}  & {EA10+/-} & {EA5+/-} \\ 
\midrule

$\mu (MPa)$ & 0.1473 & 0.079 & 0.0958 & 0.1993 & 0.3315 \\ 

$n_0kT$ & 0.041 & 0.012 & 0.0118 & 0.041 & 0.041 \\

$f (Pa)$  & 59552 & 65828 & 99780 & 60368 & 66230\\

$\kappa(MPa)$  & \multicolumn{5}{l}{6.7045} \\ 

$\dot{\gamma_0}(1/s)$  &  \multicolumn{5}{l}{0.25}\\

$\Delta G$ & \multicolumn{5}{l}{1kT}\\
 
$N_0$ & \multicolumn{5}{l}{10}\\

$\psi$ & \multicolumn{5}{l}{0.2} \\

$\delta$ &  \multicolumn{5}{l}{0.5} \\

$b(nm)$ &\multicolumn{5}{l}{3.85} \\

$\Delta G_t$ & \multicolumn{5}{l}{0.5kT}\\

$f_t(Pa)$ & \multicolumn{5}{l}{$5\times10^{6}$}\\

$k_0(1/s)$ & \multicolumn{5}{l}{1} \\

$D(m^2/s)$&\multicolumn{5}{l}{$4.1767\times10^{-20}$}\\
\bottomrule
\end{tabular}

\end{center}

\begin{figure}[H]
    \centering
    \includegraphics[width=1.2\textwidth,center,scale=0.9]{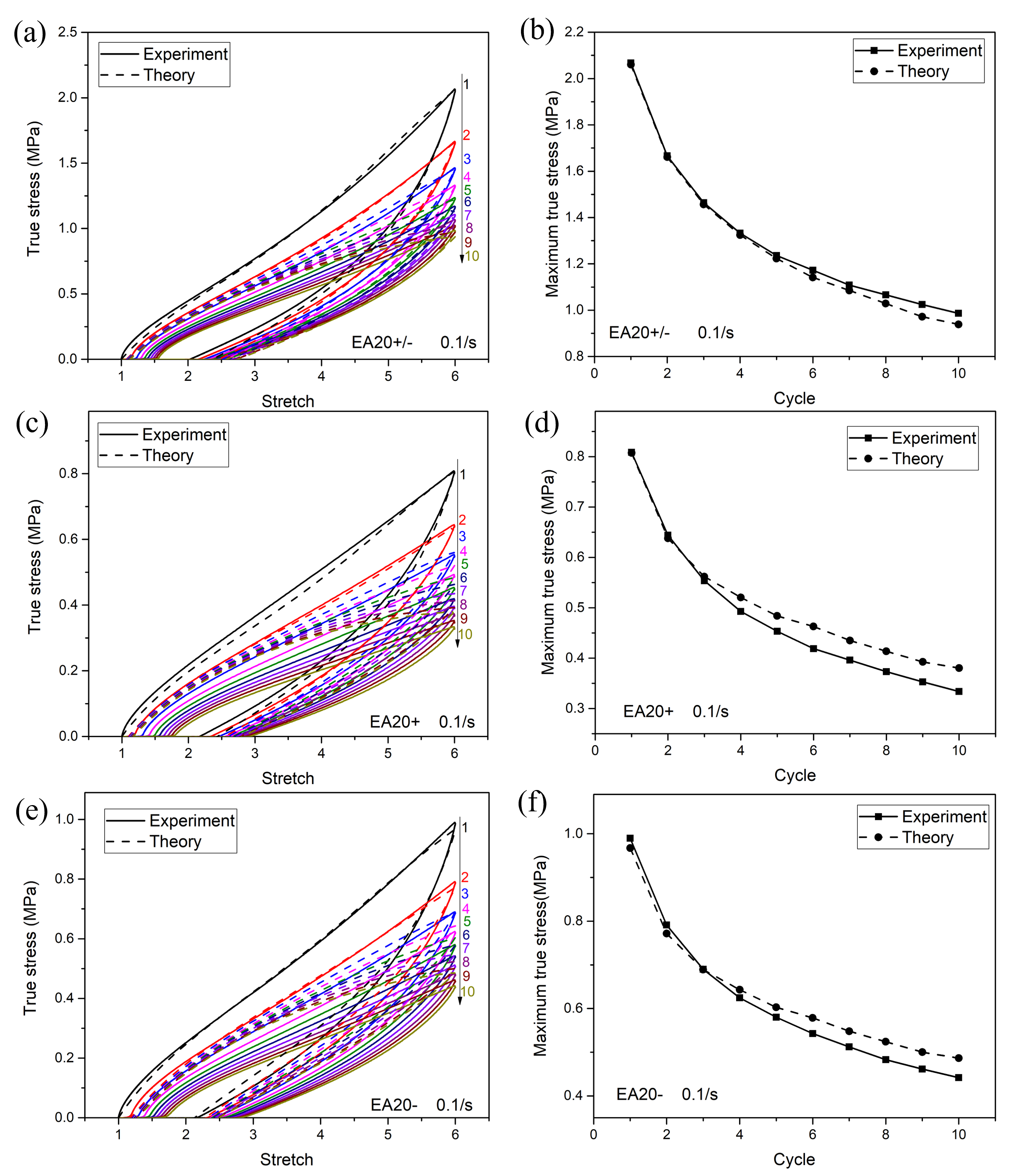}
    \caption{The fitting result of ten cycle tensile experiments at a strain rate of 0.1/s. (a) Stress-stretch for EA20+/-, and (b) the maximum true stress obtained from cyclic tensile tests. (c) Stress-stretch for EA20+, and (d) the maximum true stress obtained from cyclic tensile tests. (e) Stress-stretch for EA20-, and (f) the maximum true stress obtained from cyclic tensile tests.}
    \label{fig:cycle_relax}
    \end{figure}

Next, we use the model to capture the cyclic loading behavior of EA20+/-, EA20+, and EA20-. The true stress-stretch curve shows substantial hysteresis and has a similar shape for the individual components (EA20+ and EA20-) and the complex (EA20+/-). The complex has approximately twice the initial elastic modulus as the individual components. Over 10 cycles, the model is able to predict the main features of the experimental results including the self-recovery, a reduced elastic modulus, and reduced peak stress for EA20+/- as shown in Figure \ref{fig:cycle_relax}a - f.  Since there is no ionic crosslink for the individual component, the shear modulus and deformation resistance for plastic flow of the individual and complex should different. Deviations in conformation from having small counter ions and somewhat different charge ratios on each polymer chain are expected for the individual components compared to the complex, making it likely that the entanglement density will be different. Using the same parameters except for the parameters regarding modulus ($\mu,n_0kT, f$), we are able to capture the behavior of individual components EA20+ and EA20- at stretch rate of 0.1/s (Figure \ref{fig:cycle_relax}a - f).

\begin{figure}[H]
    \centering
    \includegraphics[width=1.2\textwidth,center]{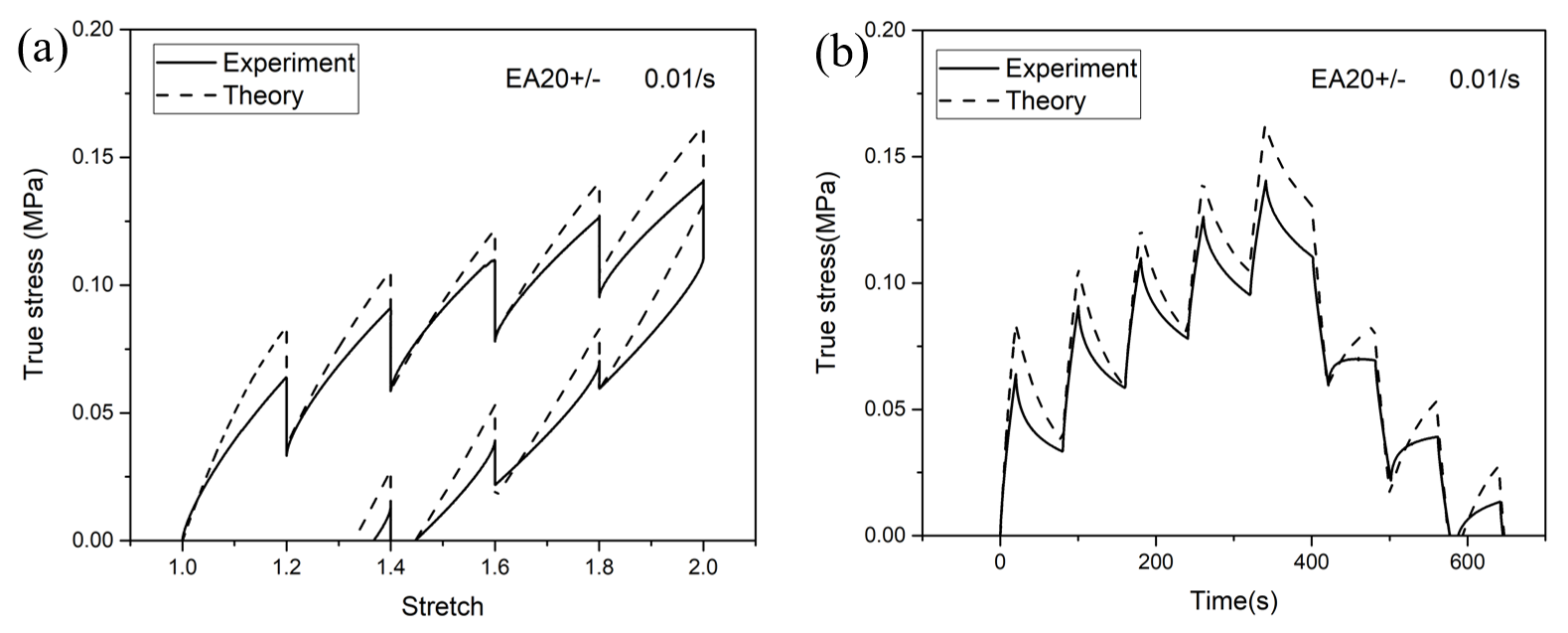}
    \caption{The fitting result for the multi-step stress relaxation experiment of EA20+/- at a strain rate of 0.01/s (a) stretch - true stress curve (b) time - true stress curve.}
    \label{fig:stress_relax}
    \end{figure}

The model is also verified by a multi-step stress-relaxation experiment. The sample is stretched at a strain rate of 0.01/s with 60 s holds at $\lambda = 0.2, 0.4, 0.6, 0.8,1.0$. The unloading process is the same as the loading.  Our model does reasonably well at predicting the decrease/increase of the true stress during each holding period during loading/unloading in Figure \ref{fig:stress_relax}a and b. However, the model overall does seem to have a longer time scale relaxation than the experiment, especially during unloading.

\begin{figure}[H]
    \centering
    \includegraphics[width=1.2\textwidth,center]{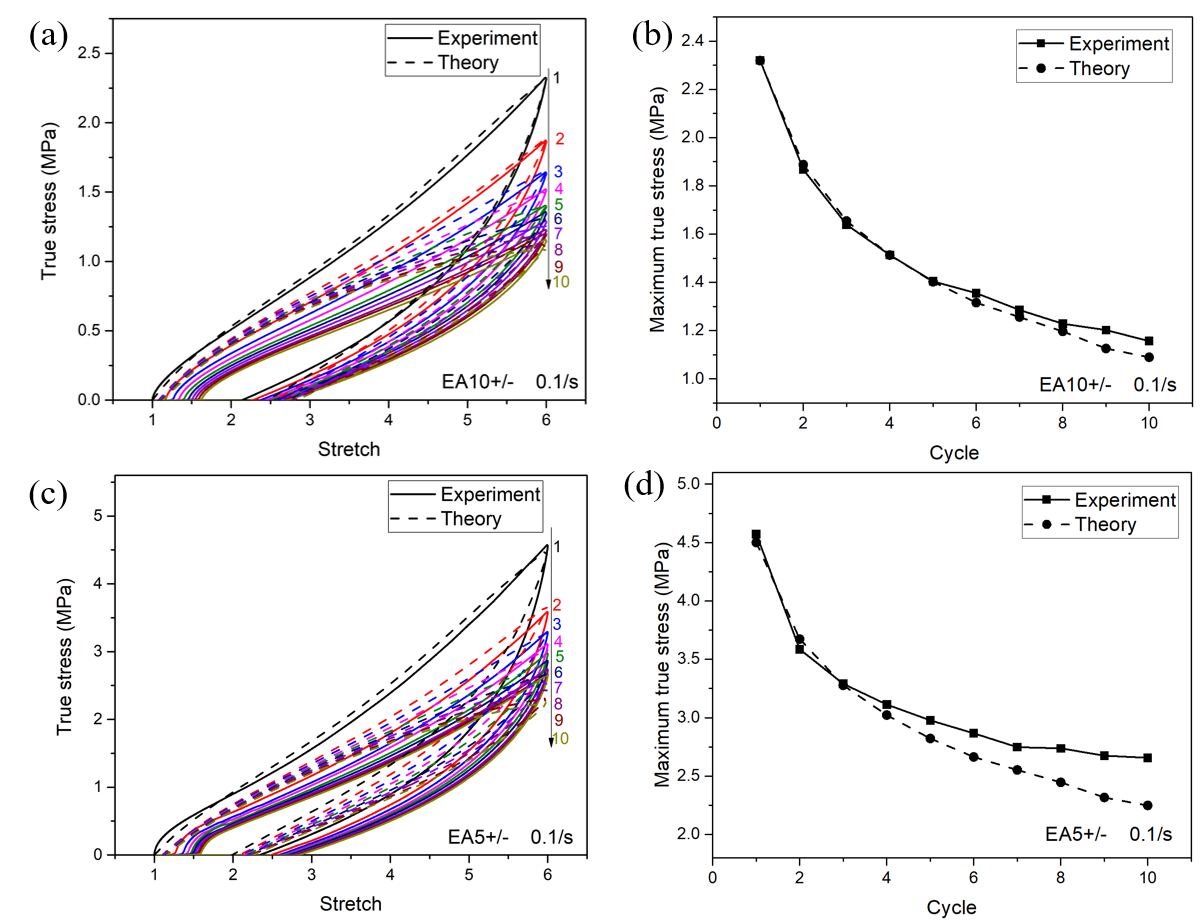}
    \caption{The fitting result of EA10+/- at a strain rate of 0.1/s for (a) ten cycle tensile experiments and (b) the maximum true stress obtained from cyclic tensile tests. The fitting result of EA5+/- at a strain rate of 0.1/s for (c) ten cycle tensile experiments, and (d) the maximum true stress obtained from cyclic tensile tests.}
    \label{fig:diff_concentration}
    \end{figure}

To help understand how our model could be used in material design, we use it to capture the mechanical behavior under multiple cycles for the bulk polyelectrolyte polymers with larger fractions of ionic components (EA10+/- and EA5+/-). The mechanical behavior for multiple cycle experiments for EA10+/- and EA5+/- are similar to EA20+/-, but with the larger ionic fraction leading to more ionic interactions, which enhances the stiffness and strength of the complexes. We therefore change the shear modulus $\mu$ and deformation resistance $f$ for EA 10+/- and EA5+/-, while keeping the other parameters the same as for EA20+/-. Fitting parameters for EA10+/- and EA5+/- are given in Table \ref{tab:parameter}. Over ten cycles, the model, with only these two parameters modified, captures the main features of the experimental results including the self-recovery, a reduced elastic modulus, and reduced peak stress for EA10+/- and EA5+/- as shown in Figure \ref{fig:diff_concentration}. $\mu$ and $f$ increase with increasing ionic fraction as expected. These values however, cannot be directly scaled by the ionic ratio.

\section{Conclusion} \label{sec:con}

In this work, we established a constitutive model to understand how ionic bonds and entanglements can be used in conjunction with each other to tailor the mechanical properties of elastomers. The model couples chain stretch, ionic bond sliding, and entanglement evolution. Entanglement evolution is driven by chain diffusion and shear stress. The model captures the overall softening shape of the uniaxial stress-strain curve, the reduction of elastic modulus with deformation and repeat cycling, the residual strain upon unloading, and self-recovery over time. The model was successfully fit to a polyelectrolyte complex. The mechanistic basis of the model was supported by the intuitive modification of material parameters when the fraction of ionic components was changed. Compared with other literature, this work firstly establishes the constitutive model that couples entanglement evolution with the ionic crosslink effects. Also, we demonstrate the theory is capable of modeling cyclic deformation (10 cycles) and stress relaxation. 

The constitutive model could be used to figure out how to design better materials. For example, we could increase stiffness by either increasing the ionic bond fraction or entanglement density. We would take Kuhn segment length dependence on monomer choice into account when making this design. We could enhance strength by increasing the strength (partial charge) of the ionic bonds. Or we could minimize cycle-to-cycle stiffness reduction by having a low Rouse friction (high chain diffusivity) so that entanglements reform quickly after being disturbed, for example by having long side chains. We also expect that our model could be applied to other dynamic bond systems such as polymers with dynamic metal-coordinated bonds, since the primary assumption we made in terms of the dynamic bonds is that they reform quickly.

Although the capabilities of the current model are salient, there are still some open questions. First, the model does not capture the large hysteresis for high strain rate. It is not clear whether there is a mechanism for ionically bonded and entanglement elastomers that we are missing/improperly modeling, or whether the experimental system deviates from being an elastomer due to time-temperature equivalence. Second, the choice of the parameters is determined by a fitting process. Although these parameters are physically reasonable, they cannot yet be predicted from the material structure. Finally, the mechanisms here are inferred from our understanding of the physics governing the material. It would be powerful to additionally have experiments that directly track the state variables, for example the rate of ionic bond sliding and the instantaneous entanglement density.








\section*{Author contributions}

\textbf{Zhongtong Wang}: Conceptualization, methodology, software, formal analysis, visualization, writing – original draft. \textbf{Hongyi Cai}: Investigation, writing-review and editing. \textbf{ Meredith N. Silberstein}: Conceptualization, methodology, project administration, supervision, writing-review and editing, and funding acquisition.

\section*{Conflicts of interest}
The authors declare no conflicts of interest.

\section*{Acknowledgements}
This work was supported by the U.S. Department of Energy, Office of Science, Basic Energy Sciences, under Award \# DE-SC0019141. This work made use of the Cornell Center for Materials Research Facilities supported by the National Science Foundation under Award Number DMR-1719875.

\bibliographystyle{elsarticle-harv} 
\biboptions{authoryear}
\bibliography{reference}




\end{sloppypar}
\end{document}